\documentclass[12pt]{article}
\usepackage{graphicx}
\usepackage[round]{natbib}
\usepackage{wrapfig}

\newcommand\lesssim{\mathrel{\hbox{\rlap{\hbox{\lower4pt\hbox{$\sim$}}}\hbox{$<$}}}}
\begin{document}

\title{Constraining the Bulk Properties of Dense Matter by Measuring Millisecond Pulsar Masses\\
A White Paper for the Astronomy and Astrophysics Decadal Survey, CFP Panel} 

\author{Paulo C. Freire, David Nice, James Lattimer, Ingrid Stairs,\\
  Zaven Arzoumanian, James Cordes, Julia Deneva {\em et al}.}

\date{\today}

\begin{titlepage}

\maketitle
\thispagestyle{empty}

\begin{center}
\includegraphics[width=250pt]{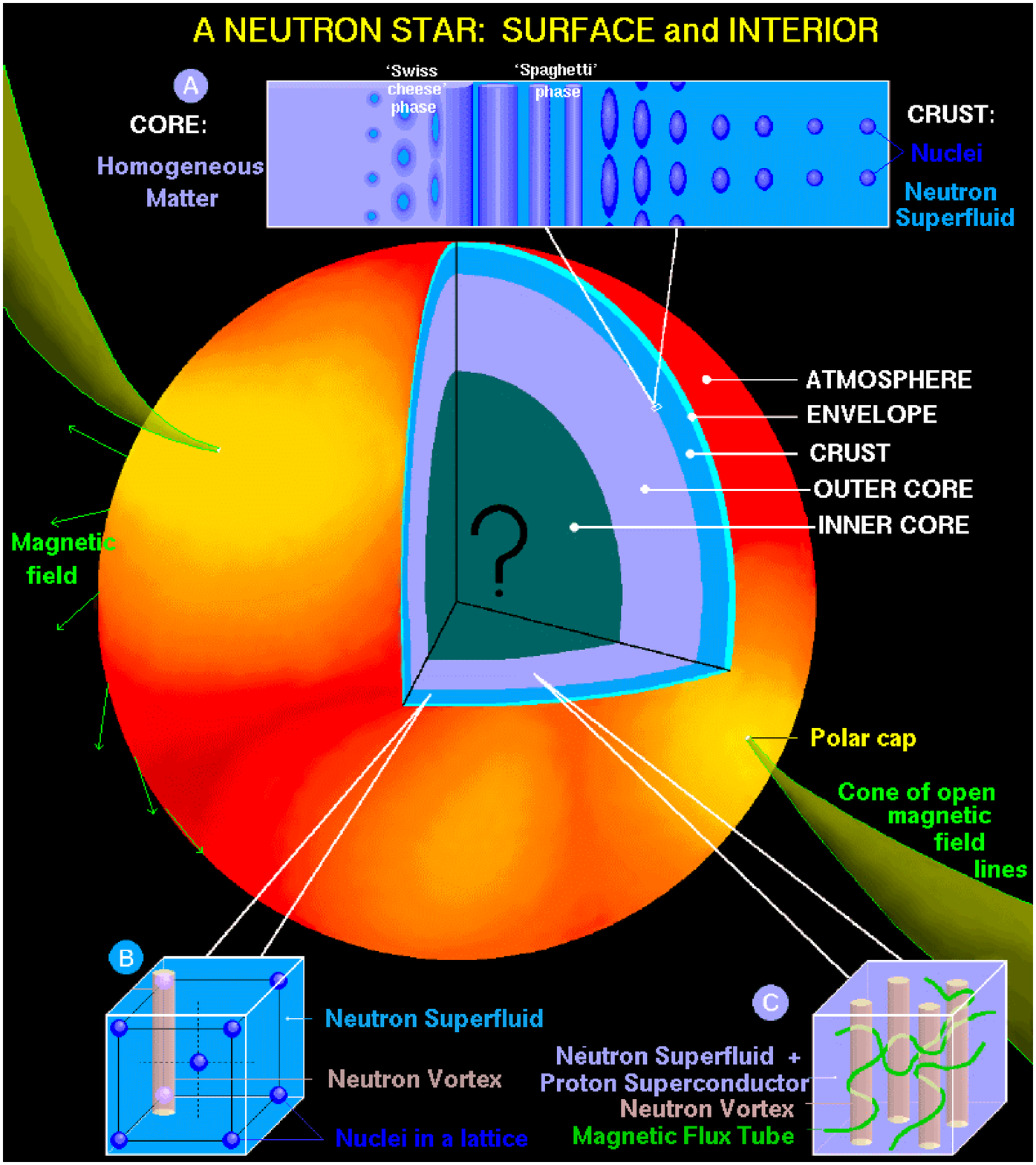}
\end{center}

{\bf The big question mark}:
More than four decades after the discovery of pulsars, the composition
of matter at their cores is still a mystery. This white paper summarizes how
recent high-precision measurements of millisecond pulsar
masses have introduced new experimental constraints on the properties
of super-dense matter, and how continued timing of intriguing new
objects, coupled with radio telescope surveys to discover more pulsars, 
might introduce
significantly more stringent constraints.  \citep[Figure from][]{lp04}.

\end{titlepage}

\section{Central Questions}\label{sec:question}

\begin{enumerate}

\item What is the nature of matter at supra-nuclear densities?

\item What is the equation of state for matter in the centers of neutron
  stars?

\item What are the upper mass limit, upper spin limit, and typical radii of neutron stars?

\item What is the distribution of masses for millisecond pulsars?

\end{enumerate}

\section{Studying the Properties of Dense Nuclear Matter}\label{sec:opportunity}
Given their small size ($R\,\sim\,10\,$km) and large mass ($M \,
\sim \, 1.4 M_{\odot}$), neutron stars contain some of the densest
matter in the Universe. 
They are unique astrophysical laboratories 
for testing theories of nuclear matter under high
pressure. 
Neutron stars may exhibit conditions and phenomena not
observed elsewhere, such as hyperon-dominated matter, deconfined quark
matter, superfluidity and superconductivity with critical temperatures
near ${10^{10}}$ kelvin, opaqueness to neutrinos, and magnetic fields
in excess of $10^{13}$ Gauss.

\begin{figure}[htp]
\begin{center}
\includegraphics[width=\textwidth]{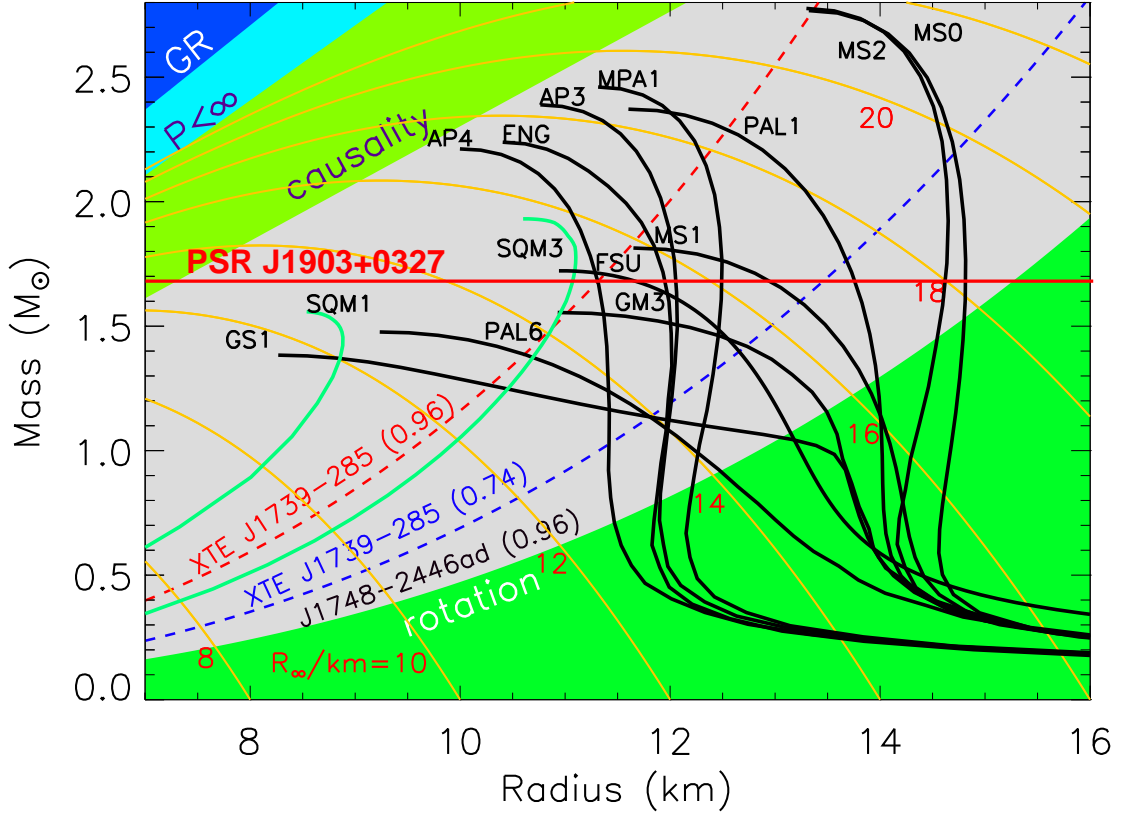}
\end{center}
\vspace*{-5ex}
\caption{%
\label{fig:EOSs} 
\small Mass-Radius relation for neutron stars. Each
black curve represents a family of neutron stars masses and
radii according to a given
equation of state. The region bounded by the Schwarzschild condition
$R < 2GM/c^2$ is excluded by general relativity, and that bounded by
$R < 3GM/c^2$ (labeled ``causality'') is excluded by requiring the speed of sound inside the
star to be smaller than the speed of light. The mass-shedding limit
for the fastest spinning radio pulsar (PSR J1748$-$2446ad, with spin
frequency 716 Hz) is labeled ``rotation''; points in the green region
below this line are not allowed for that particular pulsar.  Stricter
constraints may arise from X-ray sources like
XTE~J1739$-$285 (dashed curves calculated under different neutron
star models) if their spin frequencies are confirmed 
to be higher than that of PSR J748$-$2446ad, potentially excluding some
equations of state (such as GM3) which lie almost entirely below the
rotation curve. A recent, precise millisecond pulsar (MSP) mass
measurement (for PSR~J1903+0327) excludes the ``softest'' EOSs (red
horizontal line)
\citep[Adapted from][]{lp07}].
}
\end{figure}

The microscopic behavior of nuclear matter, encapsulated in the
equation of state (EOS), must be inferred from potentially observable,
macroscopic neutron star properties, such as their masses and radii. This
connection can be made by numerically solving the relativistic
equations of hydrostatic equilibrium (the Tolman-Oppenheimer-Volkov
equations) to calculate mass-radius ($M-R$) relations for a given EOS.
As Fig.~\ref{fig:EOSs} shows, there are large variations
in predicted radii and maximum masses for different candidate
EOSs. These reflect basic uncertainties about the behavior and
composition of matter at and above nuclear density. Some EOSs,
such as GS1, assume that large percentages of matter are in exotic states
(hyperons, deconfined quark matter), any of which produces a decrease
in the nucleonic (protonic and neutronic) degeneracy pressure at any
given density, since a larger variety of particles are present.
This means that matter is relatively more compressible; this produces
smaller maximum stellar masses.
Other EOSs, such as MSO, assume a larger fraction of nucleons and
have higher pressures for any given density.  This means that they
predict matter to be relatively incompressible and result in larger
neutron star mass limits.

\section{Constraints on Pulsar Masses and Radii}
\label{sec:context}

There are several ways in which observations of radio pulsars
constrain the properties of dense matter. 
(Considerable effort is also being put into observations of
X-ray pulsars to determine neutron star radii and masses.
This is the subject of a separate white
paper. X-ray instrumentation can also study neutron star cooling,
which is important to understand nuclear processes and the composition
of dense neutron matter well within the stellar core.)

\subsection{Constraints from pulsar spin rates}

Pulsars must be compact so that they do not shed mass due to
centrifugal forces. For a pulsar with given rotation period and mass,
this puts an upper limit on the radius. The strictest constraint comes
from the fastest-spinning pulsar. The present confirmed 
record holder is PSR~J1748$-$2446ad, which rotates at 716 Hz
\cite{hrs+06}.
This star is tantalizingly close to the centrifugal disruption spin 
frequency for several EOSs (Fig.~\ref{fig:EOSs}).  Although this spin
rate can't presently constrain the neutron star radius or the EOS,
this situation could change in the near future, as more fast-spinning
pulsars are discovered.  Especially exciting would be the discovery
of a pulsar near the centrifugal limit whose mass was measured through
other means; this would lead to limits
on the neutron star radius.

\subsection{Mass limits from double-neutron-star systems}

The best EOS constraints to date come from precise measurements of
neutron star masses.  
Constraints on masses are obtained by measuring one or more post-Keplerian (PK)
(relativistic) orbital elements in binary pulsars:
advance of periastron, varying redshift and
time dilation, orbital decay due to gravitational radiation emission,
and range and shape of Shapiro delay \citep{tay92}.  Measurement of two PK 
parameters uniquely determines
the masses of the pulsar and companion star.  (Measurement of more than
two PK parameters overconstrains the system and can be used for
tests of gravitation; this crucial application of binary pulsars is the subject of a separate white paper.)
In some sources, most notably the double pulsar,
additional constraints come from observations of the companion star.

The most precise mass limits come from double neutron star (DNS) systems,
like the Hulse-Taylor binary, PSR~B1913+16.
The pulsar in this system has a mass of
$(1.4408\,\pm\,0.0003)~{\rm M}_{\odot}$
\citep{wt03}. 
This means that if a EOS predicts a maximum
pulsar mass below $1.44~{\rm M}_{\odot}$, it is automatically excluded. The
extremely soft EOS GS1 (see Fig.~\ref{fig:EOSs}) is therefore not a
realistic description of matter at supra-nuclear densities. 

The most massive observed neutron star also sets a quantitative 
upper  limit to its central density, which then cannot be exceeded in 
any neutron star.  As successively more massive neutron stars are 
observed, the limiting central density gets smaller.  Approximately,
the limit is $\rho_{max}\le36\rho_s({\rm M}_\odot/M_{max})^2$ 
where $\rho_s=2/7\times10^{14}$ g cm$^{-3}$ is the nuclear 
saturation density.  A large enough observed mass could rule out the 
appearance of exotic phases in neutron stars, such as deconfined 
quarks \citep{lp05}.

Since the discovery of PSR~B1913+16 in 1974, at least eight other
DNS systems have been discovered.  The well-measured pulsar masses
measured range between 1.24 and $1.44~{\rm M}_{\odot}$, a surprisingly
narrow range, all near the
Chandrasekhar limit. 

\subsection{Mass limits from other pulsar binaries}
\label{sec:msp_masses}

Pulsars in neutron star-neutron star binaries discussed in the previous
section have been mildly spun up, achieving rotation periods of 10s of
milliseconds.  By contrast, pulsars in neutron star-white dwarf binaries
undergo much longer episodes of accretion and are spun up to periods
of a few milliseconds (hence the name ``millisecond pulsar,'' or MSP).  
Their fast periods lead to superior timing
precision.  However, they provide difficult targets for PK parameter
measurements because of very low eccentricities, wide orbits,
and low companion masses. 

Nevertheless, some measurements have been made in these systems.  
See, for example, figure 3 of \cite{lp07} and figure 3 of \cite{nsk08}.
In most cases, the uncertainties on the mass measurements are
frustratingly large.
In cases with unusually high eccentricity or high timing precision,
more precise measurements can be made.
In the case of PSR~J0024$-$7204H,
an MSP in a mildly eccentric ($e = 0.07$) binary system in
globular cluster (GC) 47~Tucanae, measurement of periastron advance,
together with the mass function of the system, yields $M_p < 
1.5~{\rm M}_{\odot}$ \citep{fck+03}. In the case of PSR~J1909$-$3744, a bright
MSP with a fortuitously ``edge-on'' geometry, measurement
of Shapiro delay yields
$M_p = 1.47^{+0.03}_{-0.02} M_{\odot}$ \citep{hbo06}. These
results established that at least some MSPs can be recycled with
relatively small amounts of mass, but introduced no significantly new EOS
constraints.

The situation has now changed dramatically. The most important event
has been the discovery of PSR~J1903+0327 \citep{crl+08}. This
extraordinary binary pulsar has a period similar to NS-WD
systems ($P\!=\!$2.15 ms), but is in a wide ($P_b\!=\!95$d), eccentric
orbit.  It was the first MSP discovered in the ongoing ALFA pulsar
survey. It is unique among fast MSPs in having a fairly massive
companion (either a massive white dwarf or possibly a
main-sequence star, as suggested by the spatial coincidence of a
Sun-like star with the astrometric position of the pulsar) and
absolutely extraordinary in having an orbital eccentricity of
0.44. These properties are completely at odds with all expectations
based on our present understanding of the stellar evolution in binary
systems.

In addition to being an intriguing system from a binary evolution
point of view, the unusual orbital characteristics are extremely
useful for measuring the pulsar mass.  By combining a precise 
measurement of the apsidal motion of this system
($\dot{\omega} = 86.58\pm 0.25$ arcseconds per century)
with Shapiro delay measurements, the mass has been determined
to be $1.671 \pm 0.008 {\rm M}_{\odot}$ (the $4\sigma$ uncertainty is
only $0.032 {\rm M}_{\odot}$).  Shapiro delay measurements are beginning
to over-constrain the system and confirm the unusually high mass
of this neutron star.  The precision of this measurement is
unprecedented for any MSP, and the mass is significantly higher than
any previous precise neutron star mass measurement.
If these results stand under further scrutiny, we can
exclude several soft EOSs (see red line in Fig.~\ref{fig:EOSs}).

\subsection{Millisecond Pulsars in Globular Clusters}

There has been a wealth of MSP discoveries in globular clusters
in recent years, thanks to deep observations with the Green Bank
Telescope and elsewhere; about 2/3 of the MSPs now known are located
in GCs\footnote{see http://www.naic.edu/$\sim$pfreire/GCpsr.html}.  
Because of perturbations from passing stars, or in some cases exchange
encounters, these systems can be considerably more eccentric than MSPs
in the Galactic disk, as in the case of PSR~J0024$-$7204H mentioned
above. In fact in some cases these MSP binaries are more eccentric
than any system in the Galactic disk \cite{fgri04,frg07}. The number
of highly eccentric ($e > 0.3$) MSP binaries increased from 1 known
before 2004 to 14 at present. Follow-up observations of these
pulsars over the coming years promise to yield a wealth of 
PK parameters and, therefore, measurements and constraints on
neutron star masses.


Even now, 
a statistical analysis of these measurements strongly suggests that
some of these pulsars are massive.
By considering the likelihood of different orientations of
PSRs~J1747$-$2446I and J in Terzan 5,  it has been found that there is
a 95\% probability that at least one of these objects is more massive
than $1.68\,{\rm M}_{\odot}$ \citep{rhs+05}. For PSR~B1516+02B, in M5, there
is a 95\% chance that $M_p\,>\,1.72\,{\rm M}_{\odot}$
\citep{fwvh08}. Finally, for PSR~J1748$-$2021B in NGC~6440 there is a
99\% probability of $M_p\,>\,2.0\,{\rm M}_{\odot}$ \cite{frb+08}. These
probabilities are all calculated assuming that the apsidal motion is
purely relativistic. In the case of M5B the non-identification of the
companion in HST images guarantees that the apsidal motion is
relativistic; this is not guaranteed in the case of NGC~6440B. A deep
HST observation of the environs of NGC~6440B is planned in order to
investigate the nature of its companion.


\begin{figure}
\centering
\begin{minipage}[t]{0.53\linewidth}
\includegraphics[width=1\linewidth]{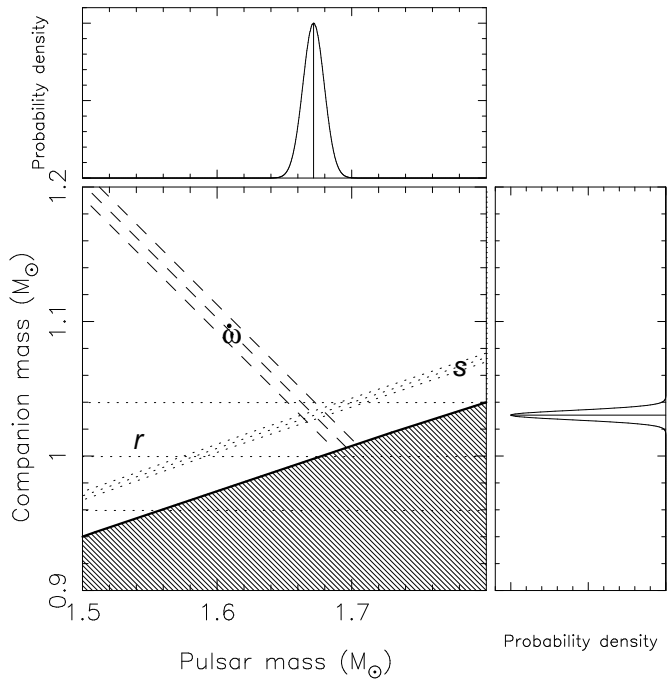}
\end{minipage}\hfill
\begin{minipage}[b]{0.45\linewidth}
\caption{\label{fig:1903+0327} \small Constraints on the masses of
PSR~J1903+0327 and its companion, based on Green Bank and Arecibo
timing. The hashed area
is excluded by knowledge of the mass function and by $| \sin i | \leq
1$, where $i$ is the orbital inclination. The remaining constraints
are derived from the rate of advance of periastron ($\dot{\omega}$)
and the ``range'' and ``shape'' ($r, s$) of the Shapiro delay, {\em
assuming} they are of general-relativistic origin. The different
constraints give a consistent measurement of the components masses,
i.e., GR passes this test. The companion is significantly lighter than
any NS measured to date, it is likely to be a
massive WD or a MS star. The pulsar is significantly more massive than
any other NS with a precise mass determination. Marginal panels, {\em
top}: probability distribution function for the mass of the pulsar,
the median is indicated with a vertical line, {\em right:} probability
distribution function for the mass of the companion. From Freire et
al. (2009).}
\end{minipage}
\end{figure}

The apsidal motions are now being measured for four more systems in
Terzan~5 and two others in M28, and it is likely that measurements
will be extended to more than a dozen GC binaries in the very near
future.  This, coupled with more
multi-wavelength observations and improved statistical methods
will allow more detailed knowledge of the underlying distribution of
MSP masses.  Further, we are now reaching the stage at which
other PK parameters, particularly $\gamma$, the time dilation
and gravitational redshift parameter, may be measurable in
some sources, particularly Terzan~5~I.

\subsection{Moment of Inertia of PSR~J0737$-$3039A}

Continued timing of the double pulsar system, J0737$-$3039, will
greatly improve the precision of its PK parameters, in particular the
measurement of the orbital decay due to the emission of gravitational
waves. Eventually, the precision of this measurement is limited by
uncertainties in the kinetic contribution to the orbital
decay. Present (Deller, Bailes and Tingay, arXiv:0902.0996) and future
interferometric studies will allow a very precise
estimate of the contribution of kinetic effects to its orbital
decay. By doing this we might be able to measure the expected spin-orbit
contribution to the $\dot{\omega}$ \citep{ds88} of PSR~J0737$-$3039A,
allowing an estimate of this pulsar's moment of inertia
\citep{kw09}. This, together with its well-determined mass,
will introduce unique and fundamental constraints on
the equation of state for dense matter \citep{ls05}. These would
complement the constraints derived from the measurement of MSP masses.


\section{Key Requirements}\label{sec:requirements}

\setlength{\leftmargini}{1em}
\begin{description}

\item[High Sensitivity Facilities]%
Pulsars are low luminosity sources, and most of the
objects that are providing the new mass limits, the GC
pulsars, are exceptionally faint.  Detection of PK parameters typically
requires consistent observation of a pulsar over a period of years.
Obtaining
any new results on these sources absolutely requires the world's most
sensitive radio telescopes, particularly the GBT and Arecibo
Observatory. These are the only instruments capable of even detecting the vast
majority of the GC pulsars---and the only ones capable of detecting
PSR J1903+0327 with the S/N required to measured the three PK
parameters we now measure for that system.
\item[Even Higher Sensitivity Facilities]%
The additional sensitivity of an instrument like the Square Kilometer
Array (SKA) would revolutionize this field, by vastly increasing the
sample of MSPs with measurable masses and greatly increasing the
number of measurable PK parameters in all sorts of systems. It will
also be able to determine the distance to PSR~J0737$-$3039 with enough
precision to allow a precise measurement of the moment of inertia of
PSR~J0737$-$3039A.
\item[Improved Radio Instrumentation]%
Precision pulsar timing requires observing the pulsar signal with high
time resolution over large radio bandwidths ($\sim$1~GHz).  The development
of the GBT and upgrades to Arecibo, have made such wide bandwidths available
for the first time, and back-end instrumentation is continually evolving
toward taking full 
advantage of these bandwidths.
This increase in bandwidth and time resolution is essential to fully
resolve the radio pulses' fine structure and increase the
sensitivity to these faint sources. 
%
\item[Continuous Long-term Observations]%
For all GC pulsars, only continued, long-term timing will allow 
measurement of the relativistic $\gamma$, which is vital for precise
measurements of pulsar masses. In some important cases, like
NGC~6440B, this will require a couple of decades. Improvements in
sensitivity and timing precision will be important to reduce these
time horizons, but the long-term nature of these projects is
essentially inevitable.

\item[New MSPs]%

Progress in using pulsar mass measurements to constrain the EOS of
neutron star matter has only been possible because of the
discovery of new objects, like PSR J1903+0327 or the GC
pulsars. This is driven by continual improvement in telescope
sensitivity, available bandwidth, computing capabilities, and pulsar
search algorithms.

Continued searches, like the ALFA pulsar survey, the 350-MHz
drift-scan survey with the GBT, the 327-MHz drift-scan survey with
Arecibo and others might discover new binary MSPs with orbital
characteristics that make their masses easy to measure. Some of
them might have masses that happen to constrain the EOS. Whatever
their masses, the more we increase the sample of MSPs with measurable
masses, the better will be our understanding of their mass distribution.

The improvements in time resolution that allow better timing precision
also allow much higher search sensitivity to fast-spinning pulsars. The
present pulsar surveys can detect sub-millisecond pulsars in Galactic
volumes that are hundreds of times larger than any previous pulsar
surveys. As discussed above, the discovery of such systems also has the
potential to fundamentally constrain the EOS.

%

\end{description}

\newcommand\apj{ApJ}
\newcommand\apjl{ApJ}

\end{document}